\documentclass[twocolumn,showpacs,preprintnumbers,amsmath,amssymb,aps, prl,setspace]{revtex4}
\newtheorem{theorem}{Theorem}
\newcommand{\Tr}{\textrm{Tr}}
\newcommand{\tr}{\footnotesize{T}}
\begin{document}


\title{\large{POVM construction: a simple recipe with applications to symmetric states}}
\author{Swarnamala Sirsi$^1$}
\author{Karthik Bharath$^2$}
\email{karthik.bharath@nottingham.ac.uk}
\author{S. P. Shilpashree$^{1,3}$}
\affiliation{$^1$Yuvaraja's College, University of Mysore, Mysore}
\affiliation{$^2$ University of Nottingham, Nottingham, U.K.}
\affiliation{$^3$K. S. School of Engineering and Management, Bangalore, India}
\date{\today}

\begin{abstract}
We propose a simple method for constructing POVMs using any set of matrices which form an orthonormal basis for the space of complex matrices. Considering the orthonormal set of irreducible spherical tensors, we examine the properties of the construction on the $N+1$-dimensional subspace of the $2^N$-dimensional Hilbert space of $N$ qubits comprising the permutationally symmetric states. Similar in spirit to Neumark's result on realization of a POVM as a projective measurement, we present a method to physically realize the constructed POVMs for symmetric states using the Clebsch--Gordan decomposition of the tensor product of irreducible representations of the rotation group. We illustrate the proposed construction on a spin-1 system, and show that it is possible to generate entangled states from separable ones. 
\end{abstract}
\maketitle

\emph{Introduction}.---Positive Operator-Valued Measures (POVMs) are used frequently in quantum information processing tasks including quantum filtering \cite{HQBB,ram}, realization of quantum communication protocols \cite{wu,wang,song,zhou}, entanglement verification \cite{RBK}, Remote State Preparation (RSP) \cite{siendong} and conclusive teleportation \cite{mor,mor1}. Their construction and physical realization are of particular interest in quantum information theory\cite{ahnert,ziman,ahnert1} with emphasis on optimal measurements through Symmetric Informationally-Complete (SIC) POVMs \cite{sic, KG} and fidelity measures \cite{DBE}. Their importance lies in the fact that they can be used to achieve certain tasks which are outside the scope of projective measurements; for example, a set of non-orthogonal states cannot be distinguished using projective measurements, but can be discriminated unambiguously using POVMs \cite{ivanovic,dieks}.  A general construction mechanism for POVMs is therefore of considerable importance. 

For $N$ qubits residing in the $2^N$-dimensional Hilbert space, the $N+1$-dimensional subspace of permutationally symmetric states corresponds to an experimentally developed and rich class \cite{elegance, usha} in quantum information processing; the class includes Bell states, Greenberger--Horne--Zeilinger (GHZ) states, W states, and $N$-qubit Dicke states, and POVMs on this class are crucial in characterization of nonlocality \cite {WM} and  multipartite entanglement \cite{ES2, ES,RM}. Given their utility in such tasks, construction of POVMs whose action always lead to symmetric post-measurement states can be particularly useful. 

The purpose then of this Letter is two-fold: to first present a simple recipe for construction of POVMs based on \emph{any} orthonormal set of basis matrices on the vector space of complex matrices; secondly, use the construction to obtain and physically realize POVMs for the permutationally symmetric  $N+1$-dimensional subspace of the $2^N$-dimensional Hilbert space in which $N$ qubits reside. The space of density matrices and POVMs is a non-linear manifold since they are required to be non-negative; however, it will be seen that the vector space structure of the space of complex matrices suffices for constructing POVMs. The nature of the construction ensures that the POVMs project states to specific subspaces spanned by a subset of the orthonormal bases matrices used;  the basis matrices need not be generators of $SU(N)$, as is usually assumed in linear representations of the density matrix. Using the representation of the density matrix with the orthonormal set of irreducible spherical tensors by Fano \cite{fano}, we present a novel method to physically realize the POVMs based on the Clebsch--Gordan decomposition of tensor products into irreducible representations of $SU(2)^{\otimes N}$, which include symmetric and anti-symmetric states. In the current context, such a procedure is similar in spirit to the construction of projective measurements from POVMs through Neumark's theorem. However, the POVMs dilated to the $2^N$-dimensional Hilbert space are not projective measurements, and generally not repeatable. Nevertheless, repeatability can be viewed in the sense that the dilated POVM \emph{always} projects onto the subspace of symmetric states. We demonstrate the physical implementation of the POVMs in a 4-dimensional 2-qubit Hilbert space, and additionally, demonstrate a mechanism to generate entangled states from initial separable states. 

\emph{General POVM construction}.---We begin with a simple observation motivating the construction. Let $x_1,\ldots, x_k$ be an orthonormal basis of a $k$-dimensional subspace $S$ of a vector space $V$ of dimension $N$, and let $X$ denote the $N \times k$ matrix whose columns are $x_1,\ldots,x_k$. Then  $XX^{\dagger}$ is Hermitian, non-negative, idempotent, and an orthogonal projection operator onto $S$, spanned by a subset of the orthonormal basis  $x_1,\ldots,x_N$ of $V$. Given any $N \times N$ matrix $A$ expressed in the basis $x_1,\ldots,x_N$, $XX^{\dagger}A$ represents the projection of $A$ onto $S$. This presents us with a potential POVM construction mechanism for projection onto subspaces of the space of density matrices coordinatized by an orthonormal basis. In order to extend this to matrix spaces, we equip the manifold of density matrices with basis matrices for the space of $N \times N$ complex matrices; while this ignores the manifold structure of the density matrix space, it provides us with a convenient mechanism for constructing POVMs. Potential matrix bases include, the generalised Gell-Mann basis, Weyl operator basis, spherical tensor basis etc. Such an approach for the Bloch vector characterization of the density matrix was adopted in \cite{BK}, wherein realization of an entanglement witness was considered. 

Let $\mathcal{M}_{N}$ denote the vector space of $N \times N$ matrices of dimension $N^2$ over the field of complex numbers $\mathbb{C}$. Define a linear map $\mathcal{V}:\mathcal{M}_{N} \to \mathcal{M}_N$,
$$\mathcal{V}(X)=\left(x_{1}^{\tr},x_{2}^{\tr},\cdots,x_{N}^{\tr}\right)^T,$$
where $x_i$ is the $i$th row of $X$ containing $N$ elements, and $v^{\tr}$ denotes the transpose of a vector $v$; the map transforms an $N \times N$ matrix into a column vector of $N^2$ elements by stacking rows into a column, providing an equivalent representation of elements of $\mathcal{M}_N$. The inverse map $\mathcal{V}^{-1}: \mathcal{M}_N \to \mathcal{M}_{N} $ is then the map which uniquely constructs an $N \times N$ complex matrix by taking the first $N$ elements of $\mathcal{V}(X)$ and setting that as the first row of $X$, and so on. Since $\mathcal{M}_{N}$ is a vector space over $\mathbb{C}$, we can choose the elementary matrix set $\{E_{i,}: i,j=1,\ldots, N\}$ of $N \times N$ matrices with 1 in the $(i,j)$th position and 0 everywhere else as an orthonormal basis; indeed, $\Tr{E_{ij}E_{i^{'}j^{'}}}=\delta_{ii^{'}}\delta_{jj^{'}}$ and $\{E_{ij}\}$ forms an orthonormal set of basis matrices for $\mathcal{M}_{N}$. Any element of $\mathcal{M}_{N}$ can be expressed as a linear combination, with complex coefficients, of the $E_{ij}$s. It is easy to verify that $E_{ij}E_{ij}^\dagger$ is Hermitian, non-negative, and satisfies
\begin{equation}\label{eij}
\displaystyle \sum_{i,j}E_{ij}E_{ij}^\dagger =N\mathbb{I}_{N}.
\end{equation}
Let us denote by $\{T_{ij}: i,j =1,\ldots,N\}$ another set of orthonormal basis matrices for $\mathcal{M}_N$. Now, using the map $\mathcal{V}$ we can express the orthonormal basis $\{E_{ij}\}$ as $\{\mathcal{V}(E_{ij}): i,j=1,\ldots,N\}$. If $U$ is an $N^2 \times N^2$ unitary matrix with $UU^\dagger=\mathbb{I}_{N^2}$, then it is possible to obtain a new orthonormal basis $\{\mathcal{V}T_{ij}\}$, where $\{\mathcal{V}T_{ij}=U\mathcal{V}(E_{ij}): i, j=1,\ldots,N\}$. Explicitly, 
\begin{equation*}
	\mathcal{V}T_{ij}=U\mathcal{V}(E_{ij})= \left(u_{ij}^{(1)},u_{ij}^{(2)},\cdots,u_{ij}^{(N)}\right)^T,
\end{equation*}
where the $N$-dimensional $u_{ij}^{(k)}$ is the $k$th sub-column vector of the $N^2$-dimensional complex vector $\mathcal{V}T_{ij}$. The inverse mapping $\mathcal{V}^{-1}$, along with unitary matrix $U$, provides us with way to obtain the $N \times N$ orthonormal basis matrices $\{T_{ij}\}$ from $\{E_{ij}\}$ as
\begin{equation*}
	T_{ij}=\mathcal{V}^{-1}(U\mathcal{V}(E_{ij}))=
	\left(\begin{array}{c}
		u_{ij}^{(1)^{\tr}}\\
		u_{ij}^{(2)^{\tr}}\\
		\vdots\\
		u_{ij}^{(N)^{\tr}}\\
	\end{array} \right).
\end{equation*}
We have thus shown how any orthonormal basis matrices for $\mathcal{M}_N$ can be constructed from the elementary basis $E_{ij}$.  Note that for any complex matrix $A$, not necessarily square, $AA^\dagger$ is always non-negative and Hermitian. From (\ref{eij}) and the fact that $UU^\dagger=\mathbb{I}_{N^2}$, it is now straightforward to verify that 
$$\displaystyle \sum_{i,j}T_{ij}T_{ij}^\dagger=\alpha \mathbb{I}_N, $$
where $\alpha$ is a constant not depending on $i$ and $j$. We can formalize the preceding discussion with the following Theorem. 
\begin{theorem}
	\label{povm}
	Let $\{T_{j}: j=1,\ldots,N^2\}$ be an orthonormal set of basis matrices for $\mathcal{M}_N$. Then the set $\{\alpha T_jT_j^\dagger\}$ satisfies the conditions of a POVM, where $\alpha$ is a constant not depending on $j$.
\end{theorem}
It is important to recognize that the subspace onto which the POVMs project a state is a vector space and is the ambient space in which the submanifold of density matrices reside. 

\emph{Spherical tensor representation of density matrix}.---A density matrix for an $N$-qubit system can be represented as
\begin{equation}\label{rho}
\rho = {1 \over (2j+1)} \sum_{k=0}^{2j}\sum_{q=-k}^k t^k_q \tau^{k^\dagger}_q,
\end{equation}
where $j=N/2$ and $\tau^k_q$ are irreducible tensor operators of rank $k$ in the $N+1$ dimensional spin space with projection $q$ along the axis of quantization in the real 3-dimensional space; here $\tau_0^0=\mathbb{I}_N$, the $N \times N$ identity operator. Here and elsewhere, orthogonality in the matrix space is always defined with respect to Trace norm or the square of the Hilbert--Schmidt norm. The $\tau^{k}_{q}$ satisfy orthogonality and symmetry relations,
\begin{equation*}
	\Tr({\tau^{k^{\dagger}}_{q}\tau^{k^{'}}_{q^{'}}})= (2j+1)\,\delta_{kk^{'}} \delta_{qq^{'}}, \quad \tau^{k^{\dagger}}_{q} = (-1)^{q}\tau^{k}_{-q},
\end{equation*}
where the normalization has been chosen so as to be in agreement with Madison convention \cite{satchler}. The Fano statistical tensors \cite{fano} or the spherical tensor parameters $t^k_q$ parametrise the density matrix $\rho$ as expectation values of $\tau^k_q$: $\Tr(\rho\tau^k_q)=t^k_q$. In other words, the $N^2-1$ spherical tensor operators $\tau^k_q$, in conjunction with the identity operator, form an orthonormal basis for the vector space of $N \times N$ complex matrices over $\mathbb{C}$ that acts on the $N+1$ dimensional spin space.  In contrast to some orthonormal basis matrices, like Gell--Mann matrices, their importance lies in the fact they can be constructed as symmetrized products of the angular momentum operators $\vec J=(J_{x}$, $J_{y}$, $J_{z})$ following  the well-known Weyl construction \cite{racah} as,
\begin{equation*}
	\tau_{q}^{k}(\vec{J}) =  \mathcal {N}_{kj}\,(\vec{J}\cdot \vec{\bf{\nabla}})^k \,r^{k} \,{Y}^{k}_{q}(\hat{r})\,,
\end{equation*} 
where 
$\mathcal {N}_{kj}$ are the normalization factors and ${Y}^{k}_{q}(\hat {r})$ are the spherical harmonics.
They possess simple transformation properties under coordinate rotations of the 3-dimensional space: for a rotation $R(\alpha,\beta,\gamma)$, where $\alpha, \beta$ and $\gamma$ are Euler angles, the parameters in the rotated coordinates, $(t^k_q)^R$, are related to the ones in the initial coordinates as
\[(t^k_q)^R= \displaystyle \sum_{q^{'}}D^k_{q^{'}q}(\alpha,\beta,\gamma)t^k_q,\]
where $D^k_{q^{'}q}(\alpha,\beta,\gamma)$ are the Wigner rotation matrices; thus the rank of the tensor is preserved under rotations. The matrix elements of the tensor operators, in the angular momentum basis, are given by
$$\langle{jm'}|\tau^{k}_{q}(\vec{J})|{jm}\rangle = [k]\,\,C(jkj;mqm'),$$
where $C(jkj;mqm')$ are the Clebsch--Gordan coefficients and $[k]=\sqrt{2k+1}$.  The tensor operators are traceless but not Hermitian, and cannot in general be identified with generators of $SU(N)$. 
\emph{POVMs for permutationally symmetric states}.---The subspace of permutationally symmetric states is spanned by the eigen states $|jm\rangle$ of angular momentum operators $J^2$ and $J_z$, where  $j=N/2$ and $m=-j,\ldots,+j$. Employing Theorem \ref{povm} in conjunction with the orthonormal set of spherical tensors from (\ref{rho}) which form a basis for the space of $N \times N$ complex matrices, we can construct POVMs for a symmetric subspace of dimension $(2j+1)$ where $j=N/2$, as
\begin{equation}\label{povm1}
E^k_q=\frac{\tau^k_q\tau^{k^\dagger}_q}{{\mathcal N}}, \qquad k=0, \ldots, 2j;  q=-k, \cdots, k,
\end{equation}
where ${\mathcal N}$ is a constant not depending on $k$ and $q$. Interestingly, using  symmetry properties of Clebsch--Gordan coefficients, it can be verified that the set $\{E^k_q\}$ contains only diagonal matrices. The symmetry relations $ \tau^{k^{\dagger}}_{q} = (-1)^{q}\tau^{k}_{-q}$ ensure that it is not possible to have $N^2-1$ distinct POVM elements; the POVMs constructed are hence neither informationally complete nor symmetric.

We now present a method which aids the physical realization of the POVM $E^k_q$ in the laboratory. Recall that $N$ qubits reside in a $2^N$-dimensional Hilbert space of which the symmetric space of dimension $N+1$ is a subspace. In this setting, Neumark's theorem states that any POVM on the $N+1$-dimensional symmetric subspace can be realized as a a projective measurement in the  $2^N$ dimensional Hilbert space. We adopt an alternative route by seeking recourse to Clebsch--Gordan decomposition of irreducible representations of $SU(2)^{\otimes N}$. From the tensor product representation of $N$ qubit state the Clebsch--Gordan decomposition leads to a direct sum of irreducible representations of which, the one with the largest dimension corresponds to the invariant subspace of symmetric states. 
In other words, we are able to obtain an orthogonal decomposition of the $2^N$-dimensional Hilbert space into invariant subspaces including symmetric and anti-symmetric states, wherein the latter does not transform under the action of $SU(2)^{\otimes N}$. In essence, the decomposition is achieved through the action of a unitary matrix with Clebsch--Gordan coefficients as elements. What is of interest in this context is that the unitary matrix decomposes the computational basis in the $2^N$-dimensional Hilbert space to a set of bases, one amongst which is the basis for the $N+1$-dimensional symmetric subspace. As an indirect consequence, we are provided with a mechanism to dilate a POVM constructed on the symmetric subspace to the $2^N$-dimensional Hilbert space. The key difference with Neumark's theorem is that the dilated POVM is not a projective measurement ensuring repeatability of measurements; instead, repeatability is to be viewed in the sense that measurements based on the dilated POVM will always result in a symmetric state. Using the constructed POVM, a measurement on an initial state $\rho^i$ results in a final state $\rho^f$ described by the density operator,
\begin{equation}
\label{after_povm}
\rho^f_{k,q}= \frac{ E^k_q \rho^i E^k_q}{Tr(E^k_q \rho^i E^k_q)} \quad. 
\end{equation}
\emph{The Spin-1 case}.---As an illustration, we now explicitly construct  the following POVMs for a spin-1 system, where  the $E^k_q$s are expressed in the basis $|1 \, 1\rangle$, $|1 \, 0 \rangle$ and $|1 \, -1 \rangle$:
{\small
\begin{align*}
E^0_0=& \left(\begin{array}{ccc} 
{1\over 9} & 0 & 0\\
0 & {1\over 9} & 0\\
0 & 0 & {1\over 9}\\
\end{array} \right),
\hspace*{1mm}
E^1_0= \left(\begin{array}{ccc} 
{1\over 6} & 0 & 0\\
0 & 0 & 0\\
0 & 0 & {1\over 6}\\
\end{array} \right),
\hspace*{1mm}
E^2_0= \left(\begin{array}{ccc} 
1 & 0 & 0\\
0 & {4 \over 18} & 0\\
0 & 0 & 1\\
\end{array} \right),
\\
E^1_1=&E^2_1= \left(\begin{array}{ccc} 
{1 \over 3} & 0 & 0\\
0 & {1 \over 3} & 0\\
0 & 0 & 0\\
\end{array} \right),
 \quad 
E^1_{-1}=E^2_{-1}= \left(\begin{array}{ccc} 
0 & 0 & 0\\
0 & {1 \over 3} & 0\\
0 & 0 & {1 \over 3}\\
\end{array} \right),
\\
&E^2_2= \left(\begin{array}{ccc} 
{1 \over 3} & 0 & 0\\
0 & 0 & 0\\
0 & 0 & 0\\
\end{array} \right),
\qquad \qquad \hspace*{2mm}
E^2_{-2} =\left(\begin{array}{ccc} 
0 & 0 & 0\\
0 & 0 & 0\\
0 & 0 & {1 \over 3}\\
\end{array} \right) .
\end{align*}
}
We can observe the degeneracy regarding the number of distinct POVM elements; we also note that $E^k_q$s do not transform like spherical tensor operators. 
We now demonstrate how starting with a POVM in a 3-dimensional symmetric subspace we can  obtain its dilation to the 4-dimensional 2-qubit Hilbert space. Consider $E^1_1$, 
in the symmetric $|1 m\rangle$ basis, $m=1,0,-1$. The relationship between $|1m\rangle$ basis and the computational basis is such that
 $|11\rangle=|\uparrow\uparrow \rangle$, 
 $|10\rangle=\frac{|\uparrow\downarrow\rangle+|\downarrow\uparrow\rangle}{\sqrt 2}$ and 
 $|1-1\rangle=|\downarrow\downarrow\rangle$.  Here the spinor in the first and second positions correspond to the first and second qubits respectively.  Let $U$ be the unitary matrix (orthogonal matrix, to be accurate, since Clebsch--Gordan coefficients are all real-valued) which transforms the computational basis  to the angular momentum basis $|11\rangle, |10\rangle,|1-1\rangle,|00\rangle$. Then, the representation of $E^1_1$ in the 2-qubit state space of dimension 4 in the computational basis $|\uparrow\uparrow\rangle, 
|\uparrow\downarrow \rangle,|\downarrow\uparrow \rangle, |\downarrow\downarrow\rangle$, is given by
\begin{equation*}
\epsilon^1_1=U(E^1_1 \oplus 0)U^\dagger,\quad \text{with} \quad 
U=\left(\begin{array}{cccc} 
1 & 0 & 0 & 0\\
0 & \frac{1}{\sqrt{2}} & \frac{1}{\sqrt{2}} & 0\\
0 & 0 & 0 & 1\\
0 &  \frac{1}{\sqrt{2}}& -\frac{1}{\sqrt{2}}& 0\\
\end{array} \right),
\end{equation*}
where $\oplus$ denotes the direct sum. Thus,
{\small
\begin{equation*}
\epsilon^1_1=\left(\begin{array}{cccc} 
1 & 0 & 0 & 0\\
0 & \frac{1}{6} & \frac{1}{6} & 0\\
0 & \frac{1}{6} & \frac{1}{6} & 0\\
0 & 0 & 0 & 0\\
\end{array} \right),\quad 
\epsilon^1_1 |\psi\rangle=\left(\begin{array}{cccc} 
1 & 0 & 0 & 0\\
0 & \frac{1}{6} & \frac{1}{6} & 0\\
0 & \frac{1}{6} & \frac{1}{6} & 0\\
0 & 0 & 0 & 0\\
\end{array} \right)
\left(\begin{array}{c}
a\\
b\\
c\\
d\\
\end{array} \right),
\end{equation*}}
and
\begin{eqnarray*}
\frac{\epsilon^1_1 |\psi\rangle}{\sqrt{\langle \psi|\epsilon^1_1 |\psi\rangle}}=\frac{\sqrt 2}{\sqrt{(2 
a^2+b^2+c^2+2bc)}}\left(\begin{array}{c}
a\\
\frac{b+c}{2}\\
\frac{b+c}{2}\\
0\\
\end{array} \right),
\end{eqnarray*}
where 
$|{\psi}\rangle=a |\uparrow\uparrow\rangle+b  |\uparrow\downarrow \rangle+c |\downarrow\uparrow \rangle+d |\downarrow\downarrow \rangle$, 
 with $|a|^2+|b|^2+|c|^2+|d|^2=1$, is the most general pure state in the 2-qubit state space. Observe that the normalized resultant state is a symmetric state given by  
 {\small
 \begin{eqnarray*}
 |\psi\rangle_{sym}=\frac{2 a|11\rangle}{\sqrt{(2 a^2+b^2+c^2+2bc)}}
 +\frac{(b+c)|10\rangle}{{\sqrt{(2 a^2+b^2+c^2+2bc)}}}\hspace*{2mm}.
 \end{eqnarray*}
}
 Therefore, $\epsilon^1_1$ projects a vector in the 4-dimensional Hilbert space onto the 3-dimensional symmetric space. Moreover,  in this example,  we show that the POVMs can be expressed in terms of Pauli spin matrices which are identified as Hamiltonians easily implemented in NMR quantum computing. If $I_1$ and $I_2$ represent the identity matrices for the qubits 1 and 2 respectively, and $\sigma_i(1)$ and  $\sigma_i(2)$, with $i=x,y,z$, the corresponding Pauli spin matrices, then $\epsilon^1_1$'s can be expressed in terms of Pauli spin 
matrices as
 \begin{align*}
\epsilon^1_1=\frac{1}{6}&\Big[(I_1 \otimes I_2)+\frac{1}{2}(\sigma_z(1) \otimes I_2)+\frac{1}{2}(I_1 \otimes \sigma_z(2))\\
&\quad+\frac{1}{2}(\sigma_x(1) \otimes \sigma_x(2)) +\frac{1}{2}(\sigma_y(1) \otimes \sigma_y(2))\Big], 
\end{align*}
where the symbol $\otimes$ denotes the direct product.
Similar calculations yield:
\begin{align*}
\epsilon^0_0&=\frac{1}{6}\Big[\frac{1}{2}(I_1 \otimes I_2)+\frac{1}{6}(\sigma_z(1) \otimes I_2)\\
&\qquad+\frac{1}{6}(I_1 \otimes 
\sigma_z(2))\nonumber -\frac{1}{6}(\sigma_x(1) \otimes \sigma_x(2))\Big] ;\\
\epsilon^1_0&=\frac{1}{12}\Big[(I_1 \otimes I_2)+(\sigma_z(1) \otimes \sigma_z(2))\\
&\quad +\frac{1}{3}(\sigma_y(1) \otimes \sigma_y(2)) 
+\frac{1}{2}(\sigma_x(1) \otimes \sigma_x(2))\Big]; 
\end{align*}

\begin{align*}
\epsilon^1_{-1}&=\frac{1}{6}\Big[(I_1 \otimes I_2) -\frac{1}{12}(\sigma_z(1) \otimes I_2)\frac{1}{12}(I_1 \otimes \sigma_z(2))\\
&\quad+\frac{1}{6}(\sigma_x(1) \otimes \sigma_x(2)) +\frac{1}{12}(\sigma_y(1) \otimes \sigma_y(2))\Big] ;\\
\epsilon^2_0&=\frac{1}{6}\Big[\frac{1}{2}(I_1 \otimes I_2)+\frac{1}{6}(\sigma_z(1) \otimes \sigma_z(2))\\
&\quad+\frac{5}{12}(\sigma_x(1) 
\otimes \sigma_x(2)) +\frac{1}{3}(\sigma_y(1) \otimes \sigma_y(2))\Big];\\
\epsilon^2_2&=\frac{1}{24}\Big[(I_1 \otimes I_2)+(\sigma_z(1) \otimes I_2)\\
&\qquad \qquad \quad+(I_1 \otimes \sigma_z(2))+(\sigma_z(1) \otimes 
\sigma_z(2))\Big];\\
\epsilon^2_{-2}&=\frac{1}{24}\Big[(I_1 \otimes I_2)-(\sigma_z(1) \otimes I_2)-(I_1 \otimes \sigma_z(2))\\
&\qquad \quad \hspace*{2mm}+
(\sigma_z(1) \otimes 
\sigma_z(2))+(\sigma_x(1) \otimes \sigma_x(2))\Big]. 
\end{align*}
 We turn our attention to the resulting state following a POVM measurement of a spin-1 density matrix . For a spin-1 system, the initial density matrix, in the representation given in (\ref{rho}), is  
given by
{\small
\begin{align*}
 \rho^i &=\frac {1}{3}\left(
    \begin{array}{ccc}
     1+\frac{3}{2} t^1_0+\frac{1}{\sqrt 2} t^2_0\quad & \frac{3}{2} (t^1_{-1}+t^2_{-1}) &\quad \sqrt{3} t^2_{-2}\\
      -\frac{3}{2} (t^1_1+t^2_1) \quad & 1-\sqrt{2} t^2_0 \quad & \frac{3}{2} (t^1_{-1}+t^2_{-1})\\
	  \sqrt{3} t^2_{2}\quad & -\frac{3}{2} (t^1_{1}+t^2_{1}) \quad & 1-\frac{3}{2} t^1_0+\frac{1}{\sqrt 2} t^2_0
	       \end{array} \right)\\
	     &:=  \left( \begin{array}{ccc}
		  \rho_{11} & \rho_{12} & \rho_{13}\\
		   \rho_{21} & \rho_{22} & \rho_{23}\\
		   \rho_{31} & \rho_{32} & \rho_{33}\\
		   \end{array} \right);			
\end{align*}
}
here $k=1$ refers to its vector polarization while $k=2$ refers to its tensor polarization. From (\ref{after_povm}) we obtain, 
\begin{equation*}
\rho^f_{10}=
\frac{1}{(\rho_{11} +  \rho_{33})}		 \left(
\begin{array}{ccc}
  \rho_{11} & 0 & \rho_{13}\\
		   0 & 0 & 0\\
		   \rho_{31} & 0 & \rho_{33}\\
		   \end{array}\right).
\end{equation*}
The only non-zero spherical tensor parameters are $t^1_0, t^2_0$ and $t^2_{\pm 2}$. Such a system can be produced  in the laboratory 
by the combined electric quadrupole and magnetic dipole field, where the direction of the magnetic field is along the $Z-$axis of 
the Principal Axis frame of the electric quadrupole field. In similar fashion,
\begin{equation*}
\rho^f_{11}=
\frac{1}{(\rho_{11} +  \rho_{22})}		 \left(
\begin{array}{ccc}
  \rho_{11} &  \rho_{12} & 0\\
		   \rho_{21} & \rho_{22} & 0\\
		   0 & 0 & 0\\
		   \end{array}\right),
		   \end{equation*}
 \begin{equation*} \rho^f_{1-1}=
\frac{1}{(\rho_{22} + \rho_{33})}		 \left(
\begin{array}{ccc}
  0 & 0 & 0\\
		   0 & \rho_{22} & \rho_{23}\\
		   0 & \rho_{32} & \rho_{33}\\
		   \end{array}\right);
\end{equation*}
the non-zero spherical parameters are $t^1_0, t^2_0, t^1_{\pm 1}$ and $t^2_{\pm 1}$. In
\begin{equation*}
\rho^f_{20}=
\frac{1}{(\rho_{11} + 16\rho_{22} + \rho_{33})}		 \left(
\begin{array}{ccc}
  \rho_{11} & 4\rho_{12} & \rho_{13}\\
		   4\rho_{21} & 16\rho_{22} & 4\rho_{23}\\
		   \rho_{31} & 4\rho_{32} & \rho_{33}\\
		   \end{array}\right),
\end{equation*}
we can see that all the spherical tensor parameters are non-zero. 
Next, we note that in the case of
\begin{equation}
\rho^f_{22}=
	 \left(
\begin{array}{ccc}
  1 & 0 & 0\\
0 & 0 & 0\\
0 & 0 & 0\\
\end{array}\right),
\qquad \rho^f_{2-2}=
	 \left(
\begin{array}{ccc}
  0 & 0 & 0\\
0 & 0 & 0\\
0 & 0 & 1\\
\end{array}\right),
\end{equation}
an arbitrary initial state collapses to a separable state after performing a measurement with $E^2_2$ and $E^2_{-2}$.

\emph{Entangled states from separable states}.---
Considering a spin-1 example, again, we demonstrate how some of the constructed POVM measurements can be used to generate entangled states from separable states. Consider $\epsilon^1_1$ acting on a separable state, 
{\small
\begin{align*}
|\psi\rangle&=		 \frac{1}{2}\left(
\begin{array}{c}
  1 \\
1 \\
1 \\
1\\
\end{array}\right) 
=
\left(\frac{|\uparrow(1)\rangle+|\downarrow(1)\rangle}{\sqrt 2}\right)\otimes \left(\frac{|\uparrow(2)\rangle+|\downarrow(2)\rangle}{\sqrt 
2}\right).
\end{align*}
}
Then, 
{\small
\[
\epsilon^1_1 |\psi\rangle=\frac{1}{3}\left(\begin{array}{cccc} 
1 & 0 & 0 & 0\\
0 & \frac{1}{2} & \frac{1}{2} & 0\\
0 & \frac{1}{2} & \frac{1}{2} & 0\\
0 & 0 & 0 & 0\\
\end{array} \right)
\left(\begin{array}{c}
\frac{1}{2}\\
\frac{1}{2}\\
\frac{1}{2}\\
\frac{1}{2}\\
\end{array} \right),\]
leading to
\begin{equation*}|\psi^f\rangle=\frac{\epsilon^1_1 |\psi\rangle}{\sqrt{\langle\psi|\epsilon^1_1 |\psi\rangle}}=\frac{1}{\sqrt 
3}\left(\begin{array}{c}
1\\
1\\
1\\
0\\
\end{array} \right),
\end{equation*}
}
where $|\psi^f\rangle$ represents the normalized state. In the computational basis, $|\psi^f\rangle$ can be represented as
\begin{equation*}
|\psi^f\rangle=\frac{1}{\sqrt 3}[|\uparrow\uparrow\rangle+|\uparrow\downarrow\rangle+|\downarrow\uparrow\rangle],
\end{equation*}
with corresponding density matrix 
\begin{align}
\label{PPT}
\rho^f=|\psi^f\rangle\langle\psi^f|
={1 \over 3}\left(\begin{array}{cccc} 
1 & 1 & 1 & 0\\
1 & 1 & 1 & 0\\
1 & 1 & 1 & 0\\
0 & 0 & 0 & 0\\
\end{array} \right)\hspace*{1mm} .
\end{align}
In order to check if $|\psi^f\rangle$ is entangled, we use the PPT criterion \cite{peres}, which offers a necessary and sufficient condition for entanglement verification in $2\times 2$ and $2\times 3$ systems.
The partially transposed matrix of $\rho^f$ in (\ref{PPT}), $\rho^f_{PPT}$, can be calculated as,
\begin{equation*}
\rho^f_{PPT}
={1 \over 3}\left(\begin{array}{cccc} 
1 & 1 & 1 & 1\\
1 & 1 & 0 & 0\\
1 & 0 & 1 & 0\\
1 & 0 & 0 & 0\\
\end{array} \right),
\end{equation*}
with eigenvalues $\lambda_1=0.872678, \lambda_2=-0.333333, \lambda_3=0.333333$, and $\lambda_4=0.127322$. Since $\lambda_2$ is negative we conclude that a separable state collapses in to an 
entangled state after a measurement is performed with $\epsilon^1_1$.


K.B. acknowledges discussions with Huiling Le about Theorem 1.

\end{document}